\documentclass[aps,showpacs,floats,amsmath,amssymb]{revtex4}
\usepackage{graphics}
\usepackage[next]{inputenc}
\usepackage[dvips]{epsfig}
\def\be{\begin{equation}}
\def\ee{\end{equation}}

\def\bi{\begin{itemize}}
\def\ei{\end{itemize}}
\def\bn{\begin{enumerate}}
\def\en{\end{enumerate}}
\def\bea{\begin{eqnarray}}
\def\eea{\end{eqnarray}}
\def\no{\nonumber}
\def\ba{\begin{array}}
\def\ea{\end{array}}
\def\bd{\begin{displaymath}}
\def\ed{\end{displaymath}}

\begin{document}
\title{phase diagram of the one dimensional
$S=\frac{1}{2}$ $XXZ$ model with Ferromagnetic nearest-neighbor
and Antiferromagnetic next-nearest neighbor interactions}

\author
{R. Jafari$^{1}$ and A. Langari$^{2}$}

\affiliation{$^1$Institute for Advanced Studies in Basic Sciences,
P.O.Box 45195-1159, Zanjan, Iran
\\$^2$Department of Physics,
Sharif University of Technology, 14588-89694, Tehran, Iran
}
\date{\today}
\begin{abstract}
We have studied the phase diagram of the one dimensional
$S=\frac{1}{2}$ $XXZ$ model with ferromagnetic nearest-neighbor
and antiferromagnetic next-nearest neighbor interactions. We have
applied the quantum renormalization group (QRG) approach to get
the stable fixed points and the running of coupling constants. The
second order QRG has been implemented to get the self similar
Hamiltonian. This model shows a rich phase diagram which consists
of different phases which possess the quantum spin-fluid and dimer
phases in addition to the classical N\'{e}el and ferromagnetic
ones. The border between different phases has been shown as a
projection onto two different planes in the phase space.

\end{abstract}
\maketitle
\section{Introduction}
There is currently much interest in quantum spin systems that
exhibit frustrations. This has been simulated in particular by
study of the magnetic properties of the cuprates which become
high- $T_{c}$ superconductors when doped. Frustrated spin systems
are known to have many interesting properties which are quite
different from the conventional magnetic systems.

The Heisenberg spin $\frac{1}{2}$ chain with nearest neighbor (NN)
and next-nearest neighbor (NNN) interactions (which is equivalent
to a zig-zag ladder) is a typical model with frustrations. In the
recent years, several interesting quasi-one-dimensional magnetic
systems have been studied experimentally
\cite{Hase,Motoyama,Coldea}. Among them, some compounds containing
$CuO$ chains with edge-sharing $CuO_{4}$ plaquette were expected
to be described by the $XXZ$ model with next-nearest neighbor
interactions. The nearest-neighbor $(Cu-Cu)$ spin interaction
changes from antiferromagnetic (AFM) to ferromagnetic (FM), as the
angle $\theta$ of the $Cu-O-Cu$ bound approaches $90^{o}$. The
next-nearest-neighbor interaction is always AFM and is not
dependent on $\theta$ \cite{Mizuno}. Several compounds with edge-
sharing chains are known, such as $Li_{2}CuO_{2}$,
$La_{6}Ca_{8}Cu_{21}O_{41}$, $Ca_{2}Y_{2}Cu_{5}O_{10}$,
$Rb_{2}Cu_{2}Mo_{3}O_{12}$, which can be considered as an ideal
model compounds with the ferromagnetic NN interactions and
antiferromagnetic NNN interactions \cite{solodovnikov,Hase-Kuroe}.

The Hamiltonian of such model on a periodic chain of $N$ sites is
\bea \label{eq1} H=\frac{J}{4}
\big\{\sum_{i=1}^{N}(\sigma_{i}^{x}\sigma_{i+1}^{x}+
\sigma_{i}^{y}\sigma_{i+1}^{y}+
\Delta~\sigma_{i}^{z}\sigma_{i+1}^{z})
+\sum_{i=1}^{N}J_{2}(\sigma_{i}^{x}\sigma_{i+2}^{x}+
\sigma_{i}^{y}\sigma_{i+2}^{y}+\delta~\sigma_{i}^{z}\sigma_{i+2}^{z})\big\},
\eea where $J>0$ and $J_{2}\geq0$ are the first and second-nearest
neighbor exchange couplings and the corresponding easy-axis
anisotropies are defined by $\Delta$ and $\Delta_{2}=J_{2}\delta$.
For $J_{2}=0$, the ground state properties are well known from the
Bethe ansatz \cite{Cloizeaux}. For positive coupling constants
($J, J_{2}, \Delta, \delta>0$), this model has been investigated
previously \cite{Nomura,Langari}. In particular, it has been shown
that a transition from a gapless state to a dimerized one takes
place at $(J_{2}=0.24, \Delta=\delta=1)$ \cite{Nomura2}. The point
$(J_{2}=\frac{1}{2}, \Delta=\delta=1)$ corresponds to the well
known Majumdar-Ghosh model where the exact ground state is
constructed from the direct products of dimers which leads to a
gapful phase \cite{Majumdar}. Relatively, less is known about the
model with the ferromagnetic NN and the antiferromagnetic NNN
interactins. Though the latter model has been a subject of many
studies \cite{Bursill,Tonegawa,Cabra,Krivnov}, the complete
picture of the phases in this model is still in investigation
\cite{Dmitriev}. It is well known that there is a critical point
$(J_{2}=\frac{1}{4},\Delta=-\delta=-1)$ where the ferromagnetic
state is unstable and the ground state is nontrivial at
$J_{2}>\frac{1}{4}$ which can be realized by different phases
\cite{Chubukov}. Moreover, the exact ground state can be
represented in the resonating valence bound state (RVB)
\cite{Bader,Hamada}. This state has been proposed as a candidate
for the spin liquid ground state \cite{Anderson}. One of the most
important and open question is the possibility of the spontaneous
dimerization of the system in the singlet phase accompanying by a
gap in the spectrum \cite{Dmitriev2}. The controversial conclusion
exists about the presence of a gap at $J_{2}>\frac{1}{4}$. It has
long been believed that the model is gapless \cite{White,Allen}
but one loop renormalization group analysis shows
\cite{Cabra,Nersesyan} that the gap opens due to a Lorentz
symmetry breaking perturbation. However, the gap has not been
checked numerically \cite{Cabra}. On the base of field theory
consideration it was proposed \cite{Itoi} that a very tiny but
finite gap exists which can  not be observed numerically.

In this paper we have considered the one dimensional anisotropic
$S=\frac{1}{2}$ Heisenberg model with ferromagnetic NN and
antiferromagnetic NNN interactions by implementing the quantum
renormalization group (QRG) method. We have calculated the
effective Hamiltonian up to the second order corrections. The
second order correction is necessary to get a self similar
Hamiltonian after each step of QRG. In this approach, we have
considered the effect of whole states of the block Hamiltonian
which are partially ignored in the first order approach. The
present scheme allows us to have the analytic RG equations, which
give a better understanding of the behavior of system by running
of coupling constants. We have succeeded in obtaining the phase
diagram in a good qualitative agreement with the numerical ones
\cite{Somma}.

We have previously studied the antiferromagnetic model
(Eq.(\ref{eq1})) for $\Delta>0$ by QRG \cite{Langari}. For
$0\leq\Delta<1$ the interplay of the two competing terms (NN and
NNN) in the presence of quantum fluctuations produces the dimer
phase for $J_{2}\geq J_{2}^{c}(\Delta, \Delta_2)$. The dimer or
spin Peierls phase has a spin gap and a broken translation
symmetry (the unit cell is doubled) in the thermodynamic limit.
However, we have determined the fluid-dimer phase transition by
using the running of couplings under RG (see Fig.3 of
Ref.\cite{Langari} or the complete phase diagram presented in
Fig.\ref{fig3} in this article). In the spin-fluid phase, the
anisotropy and next-nearest neighbor couplings are irrelevant
while in the dimer phase they run to the triple point
($\Delta_2^*=J_{2}^*\simeq0.155, \Delta^*=1$). From a quantitative
point of view at $\Delta=0$ the RG analysis gives
$J_{2}^{c}\simeq0.44$ which can be compared with the numerical
result of $J_{2}^{c}\simeq0.33$ presented in Ref.\cite{Nomura}.
The N\'{e}el phase appears just by crossing the $\Delta=1$ plane
at $\Delta_2=0$ and $J_2=0$. In the $\Delta_2=0$ plane and for
$\Delta >1$, the model will pass through a phase transition from
N\'{e}el to dimer phase for $J_2 > J_2^c(\Delta)$. The N\'{e}el
ordered is also broken by increasing the anisotropy of the NNN
interaction.

In
this paper we will complete the phase diagram of  this model for
the whole range of parameters. Moreover, we intend to consider the
$XXZ$ model with ferromagnetic NN ($J<0$) and antiferromagnetic
NNN ($J J_2 >0$) interactions which can be fulfilled by extending
the phase diagram to $J<0$ and $J J_2 >0$. If we implement a $\pi$
rotation around $z$ axis for the even sites and leave the odd
sites unchanged, the Hamiltonian (with $J<0$ and $J J_2 >0$) is transformed to the following
from \bea \label{eq2} H=\frac{J}{4}
\big\{\sum_{i=1}^{N}(\sigma_{i}^{x}\sigma_{i+1}^{x}+
\sigma_{i}^{y}\sigma_{i+1}^{y}-
\Delta~\sigma_{i}^{z}\sigma_{i+1}^{z})
+\sum_{i=1}^{N}J_{2}(\sigma_{i}^{x}\sigma_{i+2}^{x}+
\sigma_{i}^{y}\sigma_{i+2}^{y}+\delta~\sigma_{i}^{z}\sigma_{i+2}^{z})\big\};~~~
J, J_{2}, \Delta, \delta>0. \eea The QRG procedure is implemented
on the rotated Hamiltonian (Eq.(\ref{eq2})) which makes the
calculations easier. However, the phase diagram and other figures
presented in this article are based on the couplings defined in
Eq.(\ref{eq1}).

The QRG approach will be explained in the next section where the
second order effective Hamiltonian and the renormalization of the
coupling constants are obtained. In Sec. III , We will  present
the phase diagram and its characteristics where a comparison with
numerical results is done \cite{Somma}. Finally, we summarize our
results.


\section{RG equations \label{sec2}}

\begin{figure}
\begin{center}
\vspace{-2.5cm} \hspace{-2cm}
\includegraphics[width=12cm]{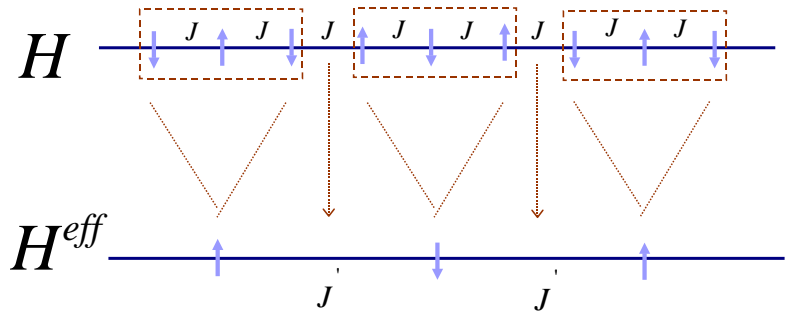}
\vspace{-2.5cm} \caption{The decomposition of chain into three site
blocks  Hamiltonian ($H^{B}$) and inter-block Hamiltonian ($H^{BB}$).}
\label{fig1}
\end{center}
\end{figure}


The main idea of the QRG method is the mode elimination or
thinning of the degrees of freedom followed by an iteration which
reduces the number of variables step by step until a more
manageable situation is reached. We have implemented the
Kadanoff's block method to do this purpose, because it is well
suited to perform analytical calculation in the lattice models and
they are conceptually easy to be extended to the higher
dimensions. In Kadanoff's method, the lattice is divided into
blocks where the Hamiltonian is exactly diagonalized. By selecting
a number of low-lying eigenstates of the blocks the full
Hamiltonian is projected into these eigenstates which gives the
effective (renormalized) Hamiltonian. The effective Hamiltonian up
to second order corrections is \cite{Langari,miguel1,miguel-2}
\be
\no H^{eff}=H^{eff}_{0}+H^{eff}_{1}+H^{eff}_{2}. \ee \bea \no
H^{eff}_{0}=P_{0}H^{B}P_{0}~~~,~~~H_{1}^{eff}=P_{0}H^{BB}P_{0}~~~,
~~~H_2^{eff}=P_{0}[H^{BB}(1-P_{0})\frac{1}{E_{0}-H^{B}}(1-P_{0})H^{BB}]P_{0}.
\eea

We have applied the mentioned scheme to the Hamiltonian defined in
Eq.(\ref{eq2}). We have considered three-site block procedure
defined in Fig.(\ref{fig1}). The block Hamiltonian ($H^{B}=\sum h_I^B $) of the three
sites, its eigenstates and eigenvalues are given in appendix A.
The three site block Hamiltonian has four doubly degenerate
eigenvalues (see appendix A). $P_{0}$ is the projection
operator to the ground state subspace which defines the RG
procedure. Due to the level crossing which occurs for the
eigenstates of the block Hamiltonian, the projection operator ($P_{0}$) can be
different depending on the coupling constants. Therefore, we must
specify the regions with the corresponding ground states. The
eigenvalues of the block Hamiltonian are labeled by $e_{0},
e_{1},e_{2}, e_{3}$ (see appendix A). In the following, we will
classify the regions where each of this states represent the ground
state. A summary of this information is given in Fig.(\ref{fig4})
of appendix A.

\subsection{Region (A): $e_{0}$ is the ground state.}
In this region the effective Hamiltonian in the first order
correction leads to the $XXZ$ chain without the NNN interaction
($J'_{2}=0$), i.e the effective Hamiltonian is not exactly similar
to the initial one. The NNN interaction is the result of the
second order correction. When the second order correction is added
to the effective Hamiltonian, the renormalized Hamiltonian, apart
from an additive constant, is similar to Eq.(\ref{eq2}) with the
renormalized couplings. Thus, the effective Hamiltonian including
the second order correction for $\Delta>0$ is:

\bea \no
H^{eff}=\frac{J'}{4}\left[\sum_{i}^{N/3}({\sigma}_{i}^{x}{\sigma}_{i+1}^{x}+{\sigma}_{i}^{y}{\sigma}_{i+1}^{y})-
\Delta'({\sigma}_{i}^{z}{\sigma}_{i+1}^{z})+\sum_{i}^{N/3}J_{2}'({\sigma}_{i}^{x}{\sigma}_{i+2}^{x}+
{\sigma}_{i}^{y}{\sigma}_{i+2}^{y})+\Delta_{2}'({\sigma}_{i}^{z}{\sigma}_{i+2}^{z})\right]
. \eea

The renormalized coupling constants are functions of the original
ones which are given in appendix B.

\subsection{Region (B): $e_{2}$ is the ground state.}

The second order effective Hamiltonian is similar to the case of
region A with different coupling constants given in appendix C. A
note is in order here, although the second order correction is
necessary to produce the NNN interaction in the effective
Hamiltonian the initial values of $J_2=0$ and $\Delta_2=0$ do not
produce NNN interactions. It is different from the RG flow
obtained in region A.

\subsection{Region (C): $e_{3}$ is the ground state.}

In this region the effective Hamiltonian to the second
order corrections leads to the Ising model
\bea
\no H^{eff}=\frac{1}{4}\left[\sum_{i}^{N/3}-
\Delta'({\sigma}_{i}^{z}{\sigma}_{i+1}^{z})\right].
\eea
Where
\bea
\no
\Delta'&=&J(\Delta+2\Delta_{2})\\
\no
&+&\frac{J^{2}}{4}\left[(\frac{1}{e_{3}-e_{0}})(\frac{1}{2+q^{2}})(1+2J_{2}q)^{2}
+(\frac{1}{e_{3}-e_{1}})(\frac{1}{2+p^{2}})(1+2J_{2}p)^{2}+(\frac{1}{e_{3}-e_{2}})(\frac{1}{2})^{2}\right]\\
\no
&+&J^{2}\left[(\frac{1}{2e_{3}-e_{1}-e_{0}})(\frac{1}{2+q^{2}})(\frac{1}{2+p^{2}})(1+2J_{2}(p+q))^{2}
+(\frac{1}{2e_{3}-e_{2}-e_{0}})(\frac{1}{2+q^{2}})(\frac{1}{2})(1+2J_{2}q)^{2}\right.\\
\no
&+&\left.(\frac{1}{2e_{3}-e_{2}-e_{1}})(\frac{1}{2+p^{2}})(\frac{1}{2})(1+2J_{2}p)^{2}\right].
\eea This simply introduces the ferromagnetic behavior. We will
discuss the phase diagram in terms of different regions defined
above in the following section.
\section{Phase diagram}
\subsection{Region (A)}

\begin{figure}
\begin{center}
\includegraphics[width=16cm]{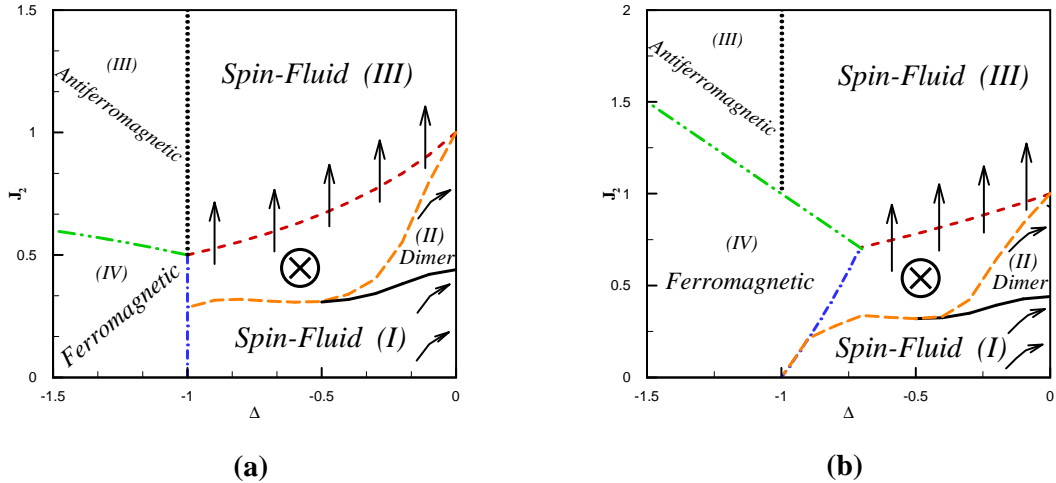}
\vspace{-2cm} \caption{The cross section of the three dimensional
phase diagram for $\Delta<0$. (a) The crossed plane is $\Delta_{2}=J_{2}\Delta$ and (b) $\Delta_{2}=0$.
} \label{fig2}
\end{center}
\end{figure}

In the $\Delta<0$ case, the RG equation shows running of $J$ to zero
which represents the renormalization of the energy scale. We have
plotted the RG flow and different phases in Fig.(\ref{fig2}). The
solid line is the boundary between dimer (II) phase and spin-fluid
(I) phase. If we start from the spin -fluid (I) phase or Dimer (II)
phase, the sign of NN anisotropy ($\Delta$) changes under RG after
few steps. However, in the dimer (II) phase the amount of NNN
coupling ($J_{2}$) is greater than 0.44 just when the anisotropy
changes sign, while in the spin-fluid (I) phase  it is less than
0.44. In other words, in dimer (II) phase the RG flow goes to the
triple point ($\Delta_2^*=J_{2}^*\simeq0.155, \Delta^*=1$) (the
filled circle in Fig.(\ref{fig3})) while it goes to
$\Delta=J_{2}=\Delta_{2}=0$ fixed point starting from the spin-fluid
(I) phase. In Fig.(\ref{fig2}(a)) and Fig.(\ref{fig2}(b)) the black
arrows show the running of couplings under RG. In the region denoted
by $\bigotimes$, though $e_{0}$ is the ground state, the behavior of
the couplings constant is not the same as the couplings in the dimer
(II) phase. In this region ($\bigotimes$) coupling constants go to
the spin-fluid (III) phase. It means that the region denoted by
$\bigotimes$ and spin-fluid (III) are a unique phase. The authors in
the reference \cite{Somma} were not able to specify the phase of
this region numerically. We denote the boundary between the
spin-fluid (III) and both dimer (II) and spin-fluid (I) phases by
long-dashed line. The dashed line behind the arrows is not a phase
boundary and just represents the two regions with differenet ground
states (see appendix A). It is known that on the
$\Delta_{2}=J_{2}\Delta$ plane and $\Delta=-1$ there is a fixed
point, namely: $\Delta=-1, J_{2}=0.25$ \cite{Bader,Hamada}. However,
our approach is not able to show this fixed point because this is on
the plane which is separated by spin-fluid (I) and ferromagnetic
phases where the level crossing occurs. Instead, close to the
$\Delta=-1$ line we found the critical value of $J_{2}^c=0.28$ which
distinguishes the spin-fluid (I) and spin-fluid (III) phases.

\subsection{Region (B)}

In this region the NNN interactions are greater than NN
interactions. The implementation of bosonization technique
combined with a meanfield analysis in the reference
\cite{Nersesyan} predicted that for $\Delta_{2}=J_{2}\Delta$ plane
and $\Delta=0$, the system might exhibit a chiral ordered phase
with gapless excitations where $J_{2}>J_{2}^c=1.26$. The predicted
critical value ($J_{2}^c$) is in well agreement with the numerical
density matrix renormalization group result \cite{Hikihara}. Our
approach shows that all coupling constants are irrelevant except
$J_{2}$ and $\Delta_{2}$. For $-1<\Delta<0$ the ratio of
$\Delta_{2}$ to $J_{2}$ goes to zero and for $\Delta<-1$ this
ratio goes to infinity. It means that in the fixed point of this
region the original spin chain decouples to two $XXZ$ chains
without next-nearest-neighbor interactions where the lattice
spacing is doubled. For $-1<\Delta<0$, the model is in the
spin-fluid (III) phase which is specified in Fig.(\ref{fig2}). The
spin-fluid (III) is different from the spin-fluid (I) phase
according to  their stable fixed points. The stable fixed point
for spin-fluid (I) is $J_2=0, \Delta_2=0$ while for spin-fluid
(III) it is $\frac{\Delta_2}{J_2}\rightarrow 0$ having $J_2,
\Delta_2 \rightarrow \infty$. Note that the level crossing of
$e_0$ and $e_2$ does not define the border between spin-fluid
(III) and dimer (II) phase. This border is defined by the running
of couplings under RG equations. For $\Delta<-1$ the model is in
the antiferromagnetic phase. In this case the model is decoupled
to two antiferromagnetic Ising chains. Thus the ground state is
long-ranged antiferromagnetic ordered
$|\uparrow\uparrow\downarrow\downarrow\uparrow\uparrow\downarrow\downarrow\cdots\rangle$.

\subsection{Region (C)}

As we pointed out in sec.\ref{sec2}-C, even after adding the second
order corrections, the original Hamiltonian is mapped to the
ferromagnetic Ising model. Ising model remains unchanged under
RG as fixed point and its properties are well known. We call this region as the
ferromagnetic phase.

\begin{figure}
\begin{center}
\includegraphics[width=16cm]{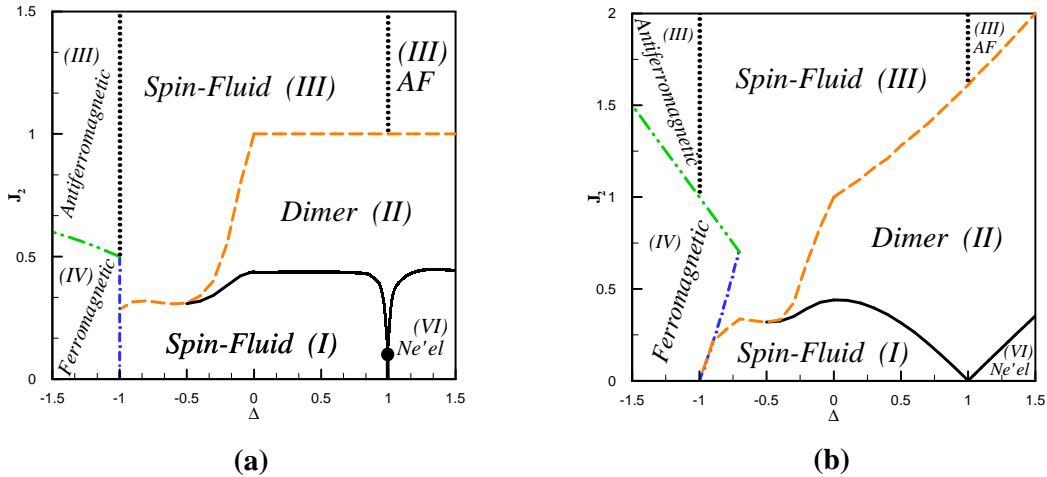}
\vspace{-2cm} \caption{The projection of the complete
phase diagram on the $\Delta{2}=J_{2}\Delta$ plane in (a)  and on the
$\Delta_{2}=0$ plane in (b).} \label{fig3}
\end{center}
\end{figure}

\section{Summary and discussions}

We mapped the one dimensional ferromagnetic NN and antiferromagnetic
NNN $S=\frac{1}{2}$ $XXZ$ model to the antiferromagnetic model in
Eq.(\ref{eq2}) with negative anisotropy. We have implemented the
second order QRG procedure to get the phase diagram of this model.
The complete phase diagram which also covers the positive anisotropy
region is presented in Fig.(\ref{fig3}). This is a cross section of
the phase diagram with $\Delta_2=J_2 \Delta$ plane in
Fig.\ref{fig3}(a) and with $\Delta_2=0$ plane in Fig.\ref{fig3}(b).
For $-0.5<\Delta<1$ (on $\Delta_{2}=J_{2}\Delta$ plane), when
$J_{2}$ is smaller than the critical value ($J_{2}^{c}$) the system
is in the gapless spin-fluid (I) phase. By contrast, for larger
value of $J_{2}>J_{2}^{c}$ the system is in the dimer phase with a
finite energy gap above the doubly degenerate ground states (the
transition is denoted by solid line). As $J_{2}$ increases, the
system exhibits a transition from the dimer phase to the spin-fluid
(III) phase which is called the gapless chiral phase in
Ref.\cite{Hikihara}. The transition is denoted by long dashed line
on the phase diagram. The transition takes place at $J_{2}^{c}=1$
for $\Delta=0$, in qualitative agreement with $J_{2}^{c}=1.26$ of
Ref.\cite{Hikihara}. For $-1<\Delta<-0.5$ (On the
$\Delta_{2}=J_{2}\Delta$ plane) at $J_{2}^{c}$ a transition occurs
from the spin-fluid(I) to spin-fluid (III) (chiral order) phases.
The QRG equations for $\Delta\gtrsim-1$ shows a critical line (long
dashed line) which separates the spin-fluid (I) and spin-fluid (III)
phases without an intermediate region. Thus, we claim that for
$J_{2}>J_2^c=0.28$ there is no gap and the model is not in the dimer
phase. The model is in ferromagnetic phase where $\Delta<-1$ and
small $J_2$. The phase transition to long-range antiferremagnetic
phase takes place at $J_{2}>J_{2}^{c}$ (dashed-dot-dot line). In the
case of $\Delta>1$, a transition from N\'{e}el (VI) phase to the dimer
phase occurs as $J_{2}$ increases (solid line). The dimer phase is
unstable by increasing $J_{2}$ further which leads to a transition
to the antiferromagnetic (AF(III)) phase (long dashed line).
However, The comparison of Fig.\ref{fig3}(a) with Fig.\ref{fig3}(b)
shows that the anisotropy of the NNN-term ($\Delta_{2}$) changes the
phase diagram significantly. From the parameters estimated for
several compounds near the isotropic limit
($\Delta=1,\Delta_{2}=J_{2}\Delta$), $La_{6}Ca_{8}Cu_{21}O_{41}$
($J_{2}^{c}=0.36$), $Li_{2}CuO_{2}$ ($J_{2}^{c}=0.62$),
$Ca_{2}Y_{2}Cu_{5}O_{10}$ ($J_{2}^{c}=2.2$) \cite{Mizuno}, our
result predict for $\Delta>-1$ that all of them are in the chiral
order phase (spin-fluid(III)) without gap, and for $\Delta<-1$,
$La_{6}Ca_{8}Cu_{21}O_{41}$ is in the ferromagnetic phase and
$Li_{2}CuO_{2}$, $Ca_{2}Y_{2}Cu_{5}O_{10}$ are in long-range
antiferromagnetic order.

\section{acknowledgment}

The authors would like to thank Prof. M. R. H. Khajehpour for careful
reading of the manuscript and 
fruitful discussions. 

\appendix

\section{The block Hamiltonian of three sites, its
eigenvectors and eigenvalues}

We have considered the three-site block (Fig.(\ref{fig1}))
with the following Hamiltonian
\bea
\no
h_{I}^{B}=\frac{J}{4}&&[(\sigma_{1,I}^{x}\sigma_{2,I}^{x}+\sigma_{2,I}^{x}\sigma_{3,I}^{x}+
\sigma_{1,I}^{y}\sigma_{2,I}^{y}+\sigma_{2,I}^{y}\sigma_{3,I}^{y})
-\Delta(\sigma_{1,I}^{z}\sigma_{2,I}^{z}+\sigma_{2,I}^{z}\sigma_{3,I}^{z})\\
\no
&&+J_{2}(\sigma_{1,I}^{x}\sigma_{3,I}^{x}+\sigma_{1,I}^{y}\sigma_{3,I}^{y})
+\Delta_{2}(\sigma_{1,I}^{z}\sigma_{3,I}^{z})],
\eea

where $\sigma_{j,I}^{\alpha}$ refers to the $\alpha$-component of
the Pauli matrix at site $j$ of the block labeled by $I$.
The exact treatment of this Hamiltonian leads to four distinct
eigenvalues which are doubly
degenerate. The ground, first, second and third excited state
energies have the following expressions in terms of the coupling
constants.

\bea \no
|\psi_{0}\rangle=\frac{1}{\sqrt{2+q^{2}}}(|\uparrow\uparrow\downarrow\rangle+
q|\uparrow\downarrow\uparrow\rangle+|\downarrow\uparrow\uparrow\rangle)~~~,~~~
|\psi_{0}'\rangle=\frac{1}{\sqrt{2+q^{2}}}(|\uparrow\downarrow\downarrow\rangle+
q|\downarrow\uparrow\downarrow\rangle+|\downarrow\downarrow\uparrow\rangle),\\
\no
e_{0}=\frac{J}{4}[~J_{2}+\Delta-\sqrt{{(J_{2}-\Delta-\Delta_{2})}^{2}+8}],~~~~~~~~~~~~~~~~~~~~~~~~~~~~~
\eea

\bea \no
|\psi_{1}\rangle=\frac{1}{\sqrt{2+p^{2}}}(|\uparrow\uparrow\downarrow\rangle+
p|\uparrow\downarrow\uparrow\rangle+|\downarrow\uparrow\uparrow\rangle)~~~,~~~
|\psi_{1}'\rangle=\frac{1}{\sqrt{2+p^{2}}}(|\uparrow\downarrow\downarrow\rangle+
p|\downarrow\uparrow\downarrow\rangle+|\downarrow\downarrow\uparrow\rangle),\\
\no
e_{1}=\frac{J}{4}[~J_{2}+\Delta+\sqrt{{(J_{2}-\Delta-\Delta_{2})}^{2}+8}],~~~~~~~~~~~~~~~~~~~~~~~~~~~~~
\eea

\bea \no
|\psi_{2}\rangle=\frac{1}{\sqrt{2}}(|\downarrow\downarrow\uparrow\rangle-
|\uparrow\downarrow\downarrow\rangle)~~~,~~~
|\psi_{2}'\rangle=\frac{1}{\sqrt{2}}(|\uparrow\uparrow\downarrow\rangle-
|\downarrow\uparrow\uparrow\rangle),\\
\no
e_{2}=\frac{-J}{4}(~2J_{2}+\Delta_{2}),~~~~~~~~~~~~~~~~~~~~~~~~~
\eea

\bea \no |\psi_{3}\rangle=|\uparrow\uparrow\uparrow\rangle~~~,~~~
|\psi_{3}'\rangle=|\downarrow\downarrow\downarrow\rangle,\\
\no e_{3}=\frac{J}{4}(\Delta_{2}-2\Delta),~~~~~~~~~ \eea

\begin{figure}
\begin{center}
\includegraphics[width=16cm]{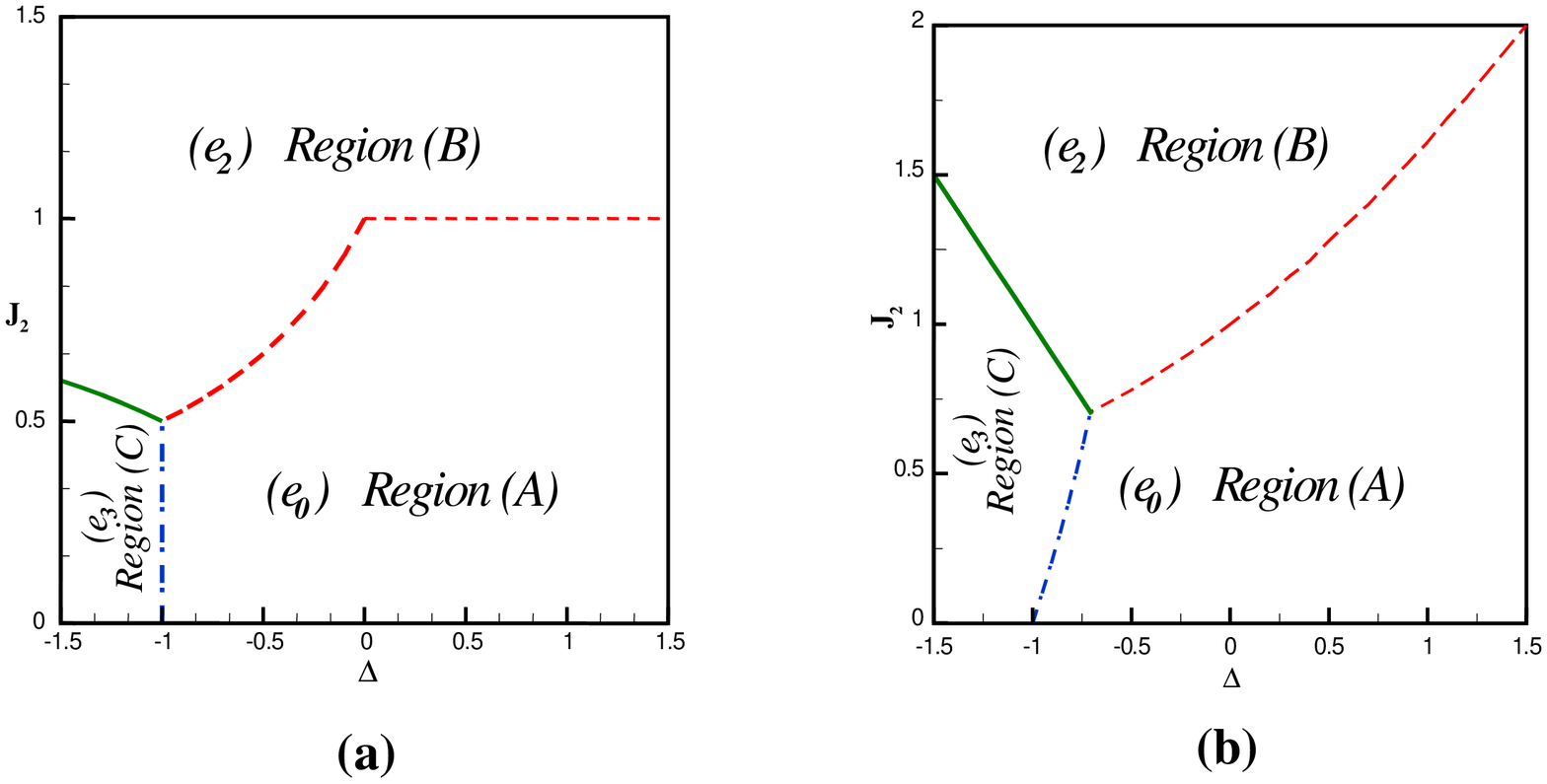}
\vspace{-1.3cm} \caption{The projection of the three dimensional
border areas on the $\Delta{2}=J_{2}\Delta$ plane (a)  and for
$\Delta_{2}=0$ plane in (b).} \label{fig4}
\end{center}
\end{figure}

where $q,p$ are

\be \no
q=\frac{-1}{2}[J_{2}-\Delta-\Delta_{2}+\sqrt{{(J_{2}-\Delta-\Delta_{2})}^{2}+8}]~~~,~~~
p=\frac{-1}{2}[J_{2}-\Delta-\Delta_{2}-\sqrt{{(J_{2}-\Delta-\Delta_{2})}^{2}+8}].
\ee

$|\uparrow\rangle$ and $|\downarrow\rangle$ are the eigenstates of
$\sigma^{z}$.

In Fig.(\ref{fig4}) we have presented the different regions where
the specified state is the ground state of the block Hamiltonian.
The border between these regions are specified as a projection to
a fixed plane. The projection to
$\Delta_{2}=J_{2}\Delta~(\Delta=\delta)$ plane is shown in
Fig.\ref{fig4}(a) and the projection to $\Delta_{2}=0$ plane is
plotted in Fig.\ref{fig4}(b).

\section{The renormalized coupling constants of the effective
Hamiltonian in region A}

\bea \no
J'&=&J(\frac{2}{2+q^{2}})^{2}(q^{2}+2J_{2}q)+\frac{J^{2}}{4}(\frac{\Delta}{e_{0}-e_{2}})(\frac{q}{2+q^{2}})^{2}\\
\no
&-&\frac{J^{2}}{4}(\frac{1}{e_{0}-e_{1}})(\frac{1}{(2+q^{2})(2+p^{2})})^{2}(p+q)
(pq)(p+q+4J_{2})(4\Delta_{2}-(\Delta+2\Delta_{2})pq)\\
\no
&-&\frac{J^{2}}{4}(\frac{4}{2e_{0}-e_{1}-e_{2}})(\frac{1}{2+q^{2}})^{2}(\frac{q}{2+p^{2}})
(p+q+2J_{2})(2\Delta_{2}-(\Delta+\Delta_{2})pq). \eea

\bea \no \Delta'&=&\left\{J(\frac{q}{2+q^{2}})^{2}(\Delta
q^{2}-2\Delta_{2}(2-q^{2}))+\frac{J^{2}}{4}
(\frac{1}{e_{0}-e_{1}})(\frac{(p+q)^{2}+4J_{2}(p+q)}{(2+q^{2})(2+p^{2})})^{2}\right.\\
\no
&+&\frac{J^{2}}{4}(\frac{1}{e_{0}-e_{2}})(\frac{q^{2}}{2(2+q^{2})})^{2}
+\frac{J^{2}}{4}(\frac{1}{e_{0}-e_{3}})(\frac{1+2J_{2}q}{2+q^{2}})^{2}\\
\no
&+&\frac{J^{2}}{4}(\frac{2}{2e_{0}-e_{1}-e_{2}})(\frac{1}{2+p^{2}})(\frac{q(~p+q+2J_{2})}{2+q^{2}})^{2}
-\frac{J^{2}}{4}(\frac{2}{2e_{0}-e_{2}-e_{3}})(\frac{q(1+qJ_{2})}{2+q^{2}})^{2}\\
\no
&-&\left.\frac{J^{2}}{4}(\frac{4}{2e_{0}-e_{1}-e_{3}})
(\frac{J_{2}(pq+q^{2}+2)+p+q}{(2+q^{2})(2+p^2)^{1/2}})^{2}\right\}/J'.
\eea

\bea \no
J'_{2}=\left\{\frac{J^{2}}{4}(\frac{2}{2+q^{2}})^3\left[\frac{(J_{2}(3q+p)+q(p+q))^{2}}{(e_{0}-e_{1})(2+p^{2})}
+\frac{(J_{2}(1+q^{2})+q)^{2}}{e_{0}-e_{3}}-\frac{(q^{2}+J_{2}q)^{2}}{2(e_{0}-e_{2})}\right]\right\}/J'.
\eea

\bea \no
\Delta'_{2}=\left\{\frac{J^{2}}{4}(\frac{2}{2+q^{2}})^{3}\left[\frac{(\Delta_{2}q(p-pq^{2}+q)
-\frac{\Delta}{2}pq^{3})^{2}}{(e_{0}-e_{1})(2+p^{2})}-
\frac{(-\Delta
q^{2}+\Delta_{2}(2-q^{2}))^{2}}{2(e_{0}-e_{2})}\right]\right\}/J'.
\eea

\section{The renormalized coupling constants of the effective Hamiltonian in region B}

\bea \no
J'&&=(\frac{J}{4})^{2}~\Delta\left[(\frac{1}{e_{2}-e_{0}})(\frac{q}{2+q^{2}})^{2}+
(\frac{1}{e_{2}-e_{1}})(\frac{p}{2+p^{2}})^{2}+
(\frac{4}{2e_{2}-e_{1}-e_{0}})(\frac{q}{2+q^{2}})(\frac{p}{2+p^{2}})\right].
 \eea

\bea \no
\Delta'&=&\left\{(\frac{J}{4})^{2}\left[(\frac{1}{e_{2}-e_{0}})(\frac{q^{2}}{2(2+q^{2})})^{2}+
(\frac{1}{e_{2}-e_{1}})(\frac{p^{2}}{2(2+p^{2})})^{2}+(\frac{1}{e_{2}-e_{3}})(\frac{1}{4})\right.\right.\\
\no
&+&\left.\left.(\frac{4}{2e_{2}-e_{1}-e_{0}})(\frac{q^{2}}{2(2+q^{2})})(\frac{p^{2}}{2(2+p^{2})})-
(\frac{2}{2e_{2}-e_{3}-e_{0}})(\frac{q^{2}}{2(2+q^{2})})\right.\right.\\
\no
&-&\left.\left.(\frac{4}{2e_{2}-e_{3}-e_{1}})(\frac{p^{2}}{2(2+p^{2})})\right]\right\}/J'.
\eea

\bea \no
J'_{2}=\left\{(\frac{JJ_{2}}{4})^{2}\left[\frac{2}{(e_{2}-e_{0})}(\frac{q^{2}}{2(2+q^{2})})
+\frac{2}{(e_{2}-e_{1})}(\frac{p^{2}}{2(2+p^{2})})+\frac{2}{(e_{2}-e_{3})}(\frac{1}{2})\right]\right\}/J'.
\eea

\bea \no
\Delta'_{2}=\left\{(\frac{J\Delta_{2}}{4})^{2}\left[\frac{2}{(e_{2}-e_{0})}(\frac{2}{2+q^{2}})
+\frac{2}{(e_{2}-e_{1})}(\frac{2}{2+p^{2}})\right]\right\}/J'.
\eea

\end{document}